\documentclass{article}
\usepackage[T1]{fontenc} 
\usepackage[utf8]{inputenc} 
\usepackage{ismir,amsmath,cite,url}
\usepackage{graphicx}
\usepackage{color,soul}

\usepackage[bookmarks=false,hidelinks]{hyperref}

\usepackage{booktabs}
\usepackage{xcolor}

\title{A Computational Analysis of Real-World DJ Mixes using Mix-To-Track Subsequence Alignment}

\multauthor
{Taejun Kim$^1$ \hspace{1cm} Minsuk Choi$^1$ \hspace{1cm} Evan Sacks$^2$} { \bfseries{Yi-Hsuan Yang$^3$ \hspace{1cm} Juhan Nam$^1$}\\
$^1$ Graduate School of Culture Technology, KAIST, South Korea\\
$^2$ 1001Tracklists, United States\\
$^3$ Research Center for IT Innovation, Academia Sincia, Taiwan\\
{\tt\small \{taejun,minsukchoi,juhan.nam\}@kaist.ac.kr, evan@1001tracklists.com, yang@citi.sinica.edu.tw}
}
\def\authorname{Taejun Kim, Minsuk Choi, Evan Sacks, Yi-Hsuan Yang, Juhan Nam}

\begin{document}

\maketitle
\begin{abstract}
A DJ mix is a sequence of music tracks concatenated seamlessly, typically rendered for audiences in a live setting by a DJ on stage. As a DJ mix is produced in a studio or the live version is recorded for music streaming services, computational methods to analyze DJ mixes, for example, extracting track information or understanding DJ techniques, have drawn research interests.
Many of previous works are, however, limited to identifying individual tracks in a mix or segmenting it, and the sizes of the datasets are usually small. In this paper, we provide an in-depth analysis of DJ music by aligning a mix to its original music tracks. We set up the subsequence alignment such that the audio features are less sensitive to the tempo or key change of the original track in a mix.
This approach provides temporally tight mix-to-track matching from which we can obtain cue-points, transition length, mix segmentation, and musical changes in DJ performance. Using 1,557 mixes from \emph{1001Tracklists} including 13,728 tracks and 20,765 transitions, we conduct the proposed analysis and show a wide range of statistics, which may elucidate the creative process of DJ music making.
\end{abstract}


\vspace{-2mm}
\section{Introduction}\label{sec:introduction}
A Disc Jockey (DJ) is a musician who plays a sequence of existing music tracks or sound sources seamlessly by manipulating the audio content based on musical elements. The outcomes can be medleys (mix), mash-ups, remixes, or even new tracks, depending on how much DJs edit the substance of the original music tracks. Among them, creating a mix is the most basic role of DJs. This involves curating music tracks and their sections to play, deciding the order, and modifying them to splice one section to another as a continuous stream. In each step, DJs consider various  
elements of the tracks 
such as tempo, key, beat, chord, rhythm, structure, energy, mood and genre. 
These days, DJs create the mix not only for a live audience but also for listeners in music streaming services.


Recently, imitating the tasks of DJ using computational methods has drawn research interests \cite{lin2014bridging,bittner2017automatic,schwarz2018heuristic, veire2018raw,Kim2017automatic,huang2018generating,huang2017djnet}.
On the other hand, efforts have been made to understand the creative process of DJ music making.
In the perspective of reverse engineering, tasks extracting useful information from real-world DJ mixes can be useful in such a pursuit.
In the literature, at least the following tasks have been studied.
(1) \textit{Track identification}~\cite{sonnleitner2016landmark,manzano2016audio,lopez2019analyzing}: identifying which tracks are played in DJ music which can be either a mix or a manipulated track.
(2) \textit{Mix segmentation}~\cite{glazyrin2014towards,scarfe2014segmentation}: finding boundaries between tracks in a DJ mix.
(3) \textit{Mix-to-track alignment}~\cite{werthen2018ground,schwarz2019methods}: aligning the original track to an audio segment in a DJ mix.
(4) \textit{Cue point extraction}~\cite{schwarz2019methods}: finding when a track starts and ends in a DJ mix.
(5) \textit{Transition unmixing}~\cite{werthen2018ground,schwarz2019methods}: explaining how DJs apply audio effects to make a seamless transition from one track to another. However, the previous studies only focused on solving the tasks usually with a small dataset and did not provide further analysis using extracted information from the tasks. For example, 
Sonnleitner et al. \cite{sonnleitner2016landmark} used 18 mixes for track identification.
Glazyrin~\cite{glazyrin2014towards} and Scarfe et al.~\cite{scarfe2014segmentation} respectively collected 103 and 339 mixes with boundary timestamps for mix segmentation.
The majority of previous studies concentrated on identification and segmentation and few studies on the other three tasks used artificially generated datasets~\cite{werthen2018ground,schwarz2019methods}.



To address the need of a large-scale study, we collected in a total of 1,557 real-world mixes and original tracks played in the mixes from \emph{1001Tracklists}, a community-based DJ music  service.\footnote{\url{https://www.1001tracklists.com}}%
The mixes include 13,728 unique tracks and 20,765 transitions.
However, tracks used in DJ mixes usually include various versions so-called ``extended mix'', ``remix'', or ``edit''.
Also, a few tracks in tracklists of the collected dataset are annotated incorrectly by users.
Therefore, an alignment algorithm is required to ensure that the collected tracks are exactly the same versions as the ones used in the mixes.
More importantly, the alignment will be a foundation for further computational analysis of DJ mixes.
With these two motivations, we set up the mix-to-track subsequence dynamic time warping (DTW)~\cite{muller2015fundamentals} such that the mix can be aligned with the original tracks in presence of possible tempo or key changes.
The warping paths from the DTW provide temporally tight mix-to-track matching from which we can obtain cue points, transition lengths, and key/tempo changes in DJ performances in a quantitative way.
To evaluate the performances of the alignment and the cue point extraction methods simultaneously, we evaluate mix segmentation performances regarding the extracted cue points as boundaries dividing two adjacent tracks in mixes, comparing them to human-annotated boundaries.
Furthermore, by observing the performance changes depending on the three different types of cue points, we analyze the human annotating policy of track boundaries.

Although DJ techniques are complicated and different depending on the characteristics of tracks, there has been common knowledge for making seamless DJ mixes.
However, to the best of our knowledge, the domain knowledge has never been addressed in the literature with statistical evidence obtained by computational analysis.
In this study, we analyze the DJ mixes using the results from the subsequence DTW mentioned above for the following hypotheses:
1) DJs tend not to change tempo and/or key of tracks much to avoid distorting the original essence of the tracks.
2) DJs make seamless transitions from one track to another considering the musical structures of tracks.
3) DJs tend to select cue points at similar positions in a single track.

The analysis is performed based on the results obtained from the subsequence alignment and provides insights statistically for tempo adjustment, key transposition, track-to-track transition lengths, and agreements of the cue points among DJs.
We hope that the proposed analysis and various statistics may elucidate the creative process of DJ music making.
The source code for the mix-to-track subsequence DTW, the cue point analysis and the mix segmentation is available at the link.\footnote{\url{https://github.com/mir-aidj/djmix-analysis/}}

\begin{table}[!t]
\centering
\scalebox{0.82}{
\begin{tabular}{lrr}
\toprule
Summary statistic & All & Matched \\
\midrule
The number of mixes                          & 1,564  & 1,557  \\
The number of unique tracks                  & 15,068 & 13,728 \\
The number of played tracks                  & 26,776 & 24,202 \\
The number of transitions                    & 24,344 & 20,765 \\
Total length of mixes (in hours)             & 1,577  & 1,570  \\
Total length of unique tracks (in hours)     & 1,038  & 913    \\
Average length of mixes (in minutes)         & 60.5   & 60.5   \\
Average length of unique tracks (in minutes) & 4.1    & 4.0    \\
Average number of played tracks in a mix     & 17.1   & 15.5   \\
Average number of transitions in a mix       & 14.5   & 12.9   \\
\bottomrule
\end{tabular}
}
\vspace{-3mm}
\caption{Statistics of the \emph{1001Tracklists} dataset.
  The original dataset size is denoted as `All' and the size after filtering as `Matched'. 
}
\label{tab:dataset}
\end{table}

\vspace{-2mm}
\section{The Dataset}\label{sec:dataset}
Our study is based on DJ music from \emph{1001Tracklists}. We obtained a collection of DJ mix metadata via direct personal communication with \emph{1001Tracklists}.
Each entry of mixes contains a list of track, boundary timestamps and genre.
It also contains web links to the audio files of the mixes and tracks. We downloaded them separately from the linked media service websites on our own. We found a small number of web links to tracks are not correct and so filtered them out by a mix-to-track alignment method automatically (see Section~\ref{sec:matchrate}). The boundary timestamps of tracks in a mix are annotated by the users of \emph{1001Tracklists}. 

\tablename~\ref{tab:dataset} summarizes statistics of the dataset. The original size of the dataset is denoted as `All' and the size after filtering as `Matched' in \tablename~\ref{tab:dataset}. Note that the number of played tracks is greater than the number of unique tracks as a track can be played in multiple mixes.
The dataset includes a variety of genres but mostly focuses on House and Trance music.
More detailed statistics of the dataset are available on the companion website.\footnote{\url{https://mir-aidj.github.io/djmix-analysis/}}

\begin{figure*}[!t]
    \centering
    \hspace{-4mm}
    \includegraphics[width=1.01\textwidth]{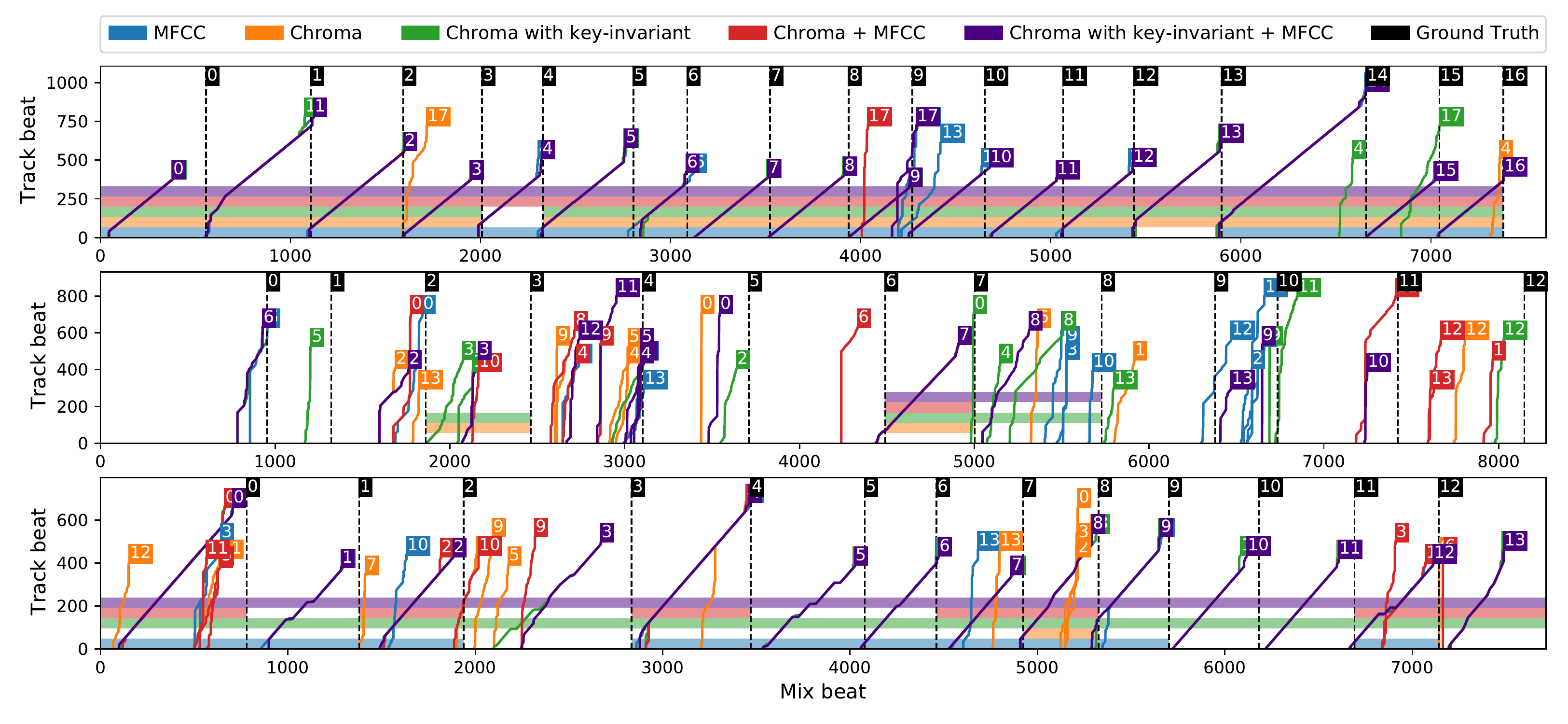}
    \vspace{-5mm}
    \caption{Visualizations of the result of a DTW-based mix-to-track subsequence alignment between a mix and the original tracks played in that mix.
    The colored solid lines show the warping paths of the alignment depending on the input feature, and whether or not applying the transposition-invariant method on the subsequence DTW.
    The tagged numbers on warping paths and ground truth boundaries indicate played and timestamped indices in the mix, respectively.
    A colored bar at the bottom of the figures is added if the alignment of the method is considered successful according to the match rate.
    (Top) A correctly matched example.
    (Middle) An unsuccessful example, due to the low sound quality of the mix.
    (Bottom) The alignment can be improved using the key-invariant chroma. Best viewed in color.
    }
    \label{fig:align}
\end{figure*}

\vspace{-2mm}
\section{Mix-To-Track Subsequence Alignment}

The objective of mix-to-track subsequence alignment is to find an optimal alignment path between a subsequence of a mix and a track used in the mix. This alignment result will be the basis of diverse DJ mix analysis concerning the cue point, track boundary, key/tempo changes and transition length. We also use it for removing non-matching tracks. This section describes the detail of computational process.

\subsection{Feature Extraction}

When DJs create a mix, they often adjust tempo and/or key of the tracks in the mix or add audio effects to them. Live mixes contain more changes in timbre and even other sound sources such as the voices from the DJ. In order to address the acoustic and musical variations between the original track and the matched subsequence in the mix, we use beat synchronous chroma and mel-frequency cepstral coefficients (MFCC). The beat synchronous feature representations enable tempo invariance and dramatically reduces the computational cost in the alignment. The aggregation of the features from the frame level to the beat level also smooths out local timbre variations. The chroma feature, on the other hand, facilitates key-invariance as circular shift of the 12-dimensional vector corresponds to key transposition. The MFCC feature captures general timbre characteristics. We used Librosa\footnote{\url{https://librosa.github.io/librosa/}} to extract the chroma and MFCC features with the default options except that the dimentionality of MFCC was set to 12 and the type of chroma was to chroma energy normalized statistics (CENS)~\cite{muller2011chroma}.

\subsection{Key-Invariant Subsequence DTW}
\label{sec:dtw}
We compute the alignment by applying subsequence DTW to the beat synchronous features~\cite{muller2015fundamentals}. We used an implementation from Librosa, adopting the transposition-invariant approach from \cite{muller2007transposition}. Specifically, we calculated 12 versions of chroma features by performing all possible circular shifts on the original track side and select the one with the lowest matching cost in the subsequence DTW.
This result returns not only the optimal alignment path but also the key transposition value of the original track. 


\figurename~\ref{fig:align} shows three examples of the alignment results when different combinations of features (MFCC, chroma, and key-invariant chroma) are used.
When the alignment path of the subsequence satisfies a match rate (described in Section \ref{sec:matchrate}), we put a color strip corresponding to each feature in the bottom of the figure.
Since we use beat synchronous representations for them, the warping paths become diagonal with a slope of one if a mix and a track are successfully aligned.
The top panel in the figure shows an successfully aligned example for the most of tracks and features where all warping paths have straight diagonal paths.\footnote{\url{https://1001.tl/14jltnct}}
The middle panel shows a failing example because sounds from crowds are also recorded in the mix.\footnote{\url{https://1001.tl/15fulzc1}}
The bottom panel shows a example where chroma with circular shift distinctively works better others as the DJ frequently uses key transposition on the mix.\footnote{\url{https://1001.tl/bcx2z0t}}



\subsection{Filtering Using Match Rates}
\label{sec:matchrate}
As stated above, we can measure the quality of the alignment from the warping path. Ideally, when every single move on the path is diagonal, that is, one beat at a time for both track and mix axis, we will obtain a perfect straight diagonal line. However, the acoustic and musical changes deform the path. We define the ratio of the diagonal moves in a mix (one move per beat) as the \textit{match rate} and use it for filtering out incorrectly annotated tracks. We experimentally chose 0.4 as a threshold. The size of the dataset after the filtering is denoted as ``Matched'' in 
\tablename~\ref{tab:dataset}. We only use the matched tracks for the  analysis in this paper.


\begin{figure}[t]
\centering
\hspace{-.4cm}
\includegraphics[width=\columnwidth]{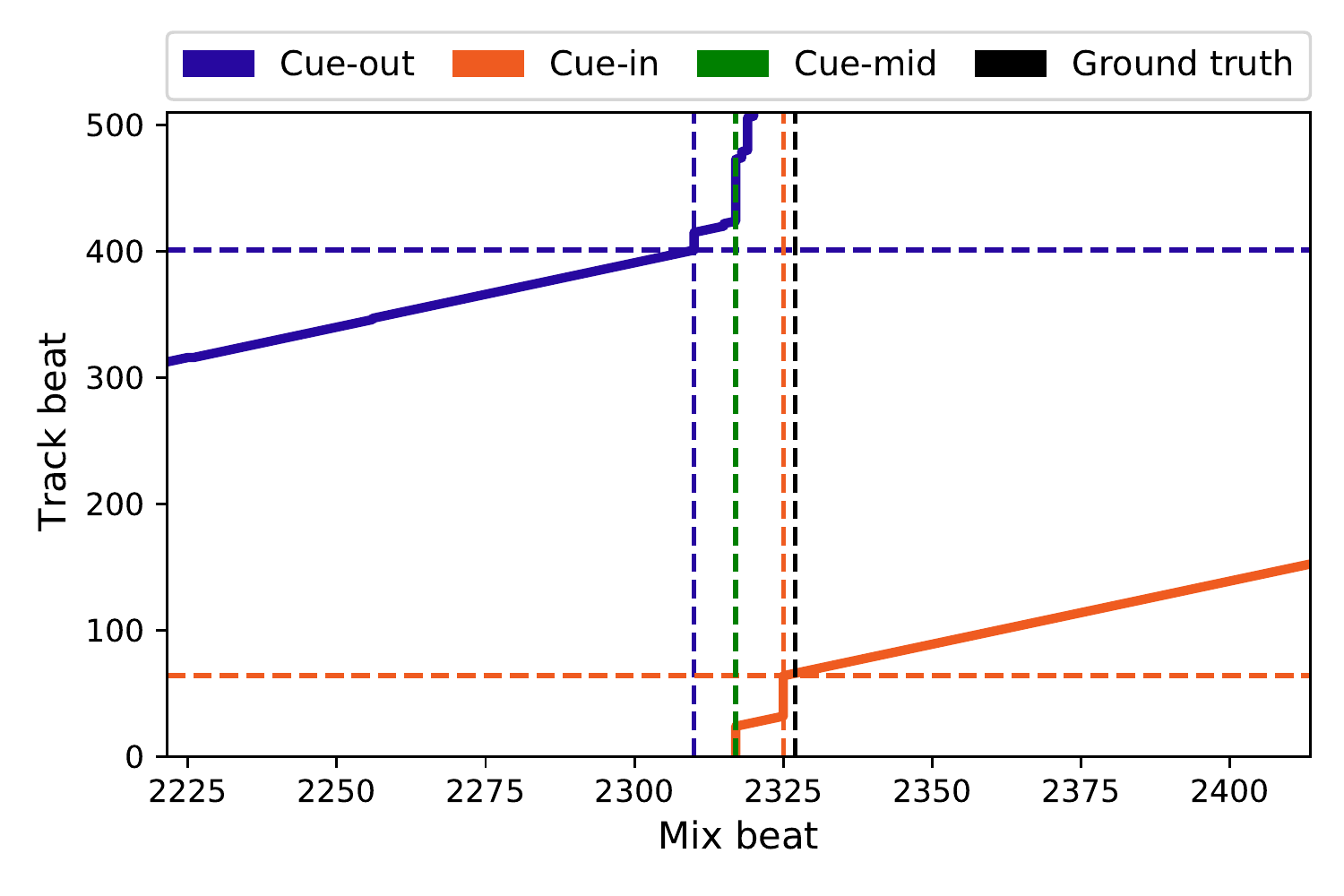}
\vspace{-5mm}
\caption{
  A zoomed-in view of a visualization of mix-to-track subsequence alignment explaining the three types of extracted cue points. The two solid lines indicate warping paths representing alignment between the mix and tracks.
  The vertical colored dotted lines represent the extracted cue points on the mix and the horizontal dotted lines represent the points on each track.
  The vertical black dotted line is a human-annotated ground truth boundary between the two tracks.
  The solid lines are the fourth and fifth warping paths from the top of \figurename~\ref{fig:align}. Best viewed in color.
}
\label{fig:cue}
\end{figure}

\vspace{-2mm}
\section{Cue Point Extraction}

Cue points are timestamps in a track that indicate where to start and end the track in a mix.
Determining the cue points of played tracks is an essential task of DJ mixing.
This section describes extracting cue points using the warping paths obtained from the aforementioned mix-to-track subsequence alignment.

\subsection{Term Definitions}
We first define terms related to cue points. In the context of the track-to-track transition, a \textit{cue-out} point is a timestamp that the previous track starts fading out and the next track starts fading in, and a \textit{cue-in} point is when the previous track is fully faded out and only the next track is being played. The \textit{transition} region is defined as the time interval from the cue-out point of the previous track to the cue-in point of the next track. Additionally, we define a \textit{cue-mid} point as the middle of a transition, which can technically be considered as a boundary of the transition.



\subsection{Methods}

The mix-to-track alignment results naturally yield cue points of matched tracks.
\figurename~\ref{fig:cue} shows an example of extracted cue points (a zoomed-in view of the top figure in \figurename~\ref{fig:align}).
The two alignment paths drift from the diagonal lines in the transition region (between 2310 and 2324 in mix beat) because the two tracks cross-fades. Based on this observation, we detect the cue-out point of the previous track by finding the last beat where preceding 32 beats have diagonal moves in the alignment path. Likewise, we detect the cue-in point of the next track by finding the first beat where succeeding 32 beats have diagonal moves in the alignment path.

\begin{figure*}[!t]
    \centering
    \hspace{-3mm}
    \includegraphics[width=0.85\textwidth]{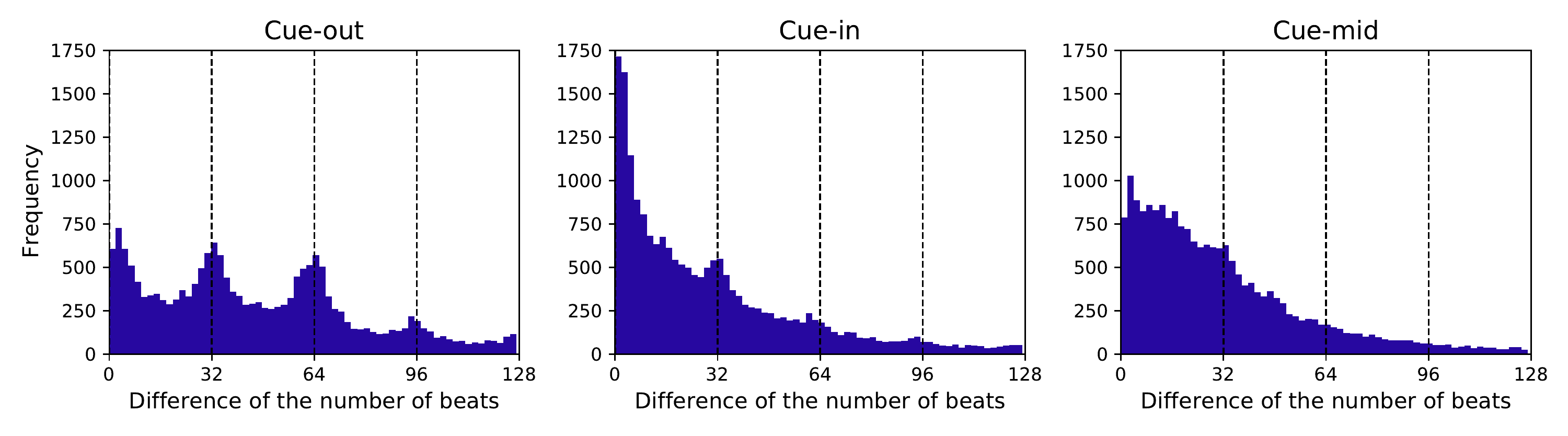}
    \vspace{-5mm}
    \caption{Histograms of distances to ground truth boundaries in the number of beats depending on the type of the cue point. The dotted lines are plotted at every 32 beats which is usually considered as a phrase in the context of dance music.}
    \label{fig:align_beatdiff}
\end{figure*}
\begin{table*}[!t]
\begin{center}
\scalebox{0.9}{
\begin{tabular}{lrrrrcrrr}
\toprule
                  & \multicolumn{4}{c}{Median time difference (in seconds)}  & & \multicolumn{3}{c}{Cue-in hit rate} \\
\cmidrule{2-5} \cmidrule{7-9}
Feature           & Cue-out & Cue-in & Cue-mid & Cue-best$^\dagger$ & & 15 sec        & 30 sec        & 60 sec \\
\hline
\midrule
MFCC              & 27.92   & 14.27  & 13.55        & 5.340 & & 0.5187     & 0.7591     & 0.9023 \\
\hline
Chroma            & 23.85   & 11.80  & 12.33        & \textbf{4.230} & & 0.5837     & 0.7973     & 0.9286 \\
Chroma with key-invariant & 23.87   & 11.77  & 12.37        & 4.240 & & 0.5843     & 0.7968     & 0.9282 \\
\hline
Chroma + MFCC       & 23.41   & 11.48  & \textbf{12.16} & 4.380 & & 0.5866 & 0.8035 & 0.9284 \\
Chroma with key-invariant + MFCC & \textbf{23.38} & \textbf{11.40} & \textbf{12.16} & 4.380 & & \textbf{0.5881} & \textbf{0.8040} & \textbf{0.9288} \\
\bottomrule
\end{tabular}
}
\end{center}
\vspace{-4mm}
\caption{
  Mix segmentation performances depending on the type of cue point and the input feature used to obtain the warping paths.
  Median time differences between cue points and ground truths are shown on the left side and hit rates of cue-in points with thresholds in seconds are shown on the right side.
  ``Key-invariant" indicates applying the key transposition-invariant method for the DTW.
  The best score of each criteria is shown in \textbf{bold}.
  $\dagger$ indicates the scores are computed using the best score among the three cue types.
}
\label{tab:hit}
\end{table*}

\begin{table}[t]
\centering
\scalebox{0.9}{
\begin{tabular}{ccc}
\toprule
Cue-out & Cue-in & Cue-mid \\
\midrule
6,151 (30\%) & 10,844 (52\%) & 3,770 (18\%) \\
\bottomrule
\end{tabular}
}
\caption{The number of ground truth boundary timestamps closest to the type of cue point.}
\label{tab:policy}
\end{table}

\section{Mix Segmentation}
The goal of mix segmentation is to divide a continuous DJ mix into individual tracks, which can enhance the listening experience and can be a foundation of further analysis or learning of DJ mixes.
Since DJs make seamless transitions, it is difficult to notice that a track is fading in or out.
To quantitatively measure how difficult it is, a study analyzed how accurate humans are at creating the boundary timestamps and found that the standard deviation of the human disagreement for track boundaries in mixes is about 9 seconds, which implies it is difficult to find the optimal boundaries even for humans~\cite{scarfe2014segmentation}. Furthermore, the ambiguous definition of the boundary and long lengths of transitions makes it difficult to annotate the boundary timestamps~\cite{sonnleitner2016landmark}. 

\subsection{Cue Point based Estimation}
Given the extracted cue point so far, we can estimate the track boundaries with three possible choices. The first is the position that the next track fully appears (cue-in point), the second is the position that previous track starts to disappear (cue-out point), and the last is the middle of the transition (cue-mid point). By comparing each of them with human-annotated boundary timestamps, we can measure which type of cue point humans tend to consider as a boundary. 

\figurename~\ref{fig:align_beatdiff} shows three histograms where each of them is computed from the differences between human-annotated boundary timestamps and one of the cue point types in beat unit. The overall trend shows that the distribution of cue-in point is mostly skewed towards zero. Interestingly, the distribution of cue-out point has more distinctive peaks around every 32 beat than the distribution of cue-in point. Considering the histogram of the transition length has peaks at every 32 beat as shown in \figurename~\ref{fig:cue_translength}, this reflects that human annotators tend to label cue-in points as a boundary compared to cue-out (note that the transition length is computed by subtracting the cue-in point from the cue-out point). On the other hand, the distribution of cue-mid point has a gradually decreasing curve without peaks. While this distribution looks like having better estimates than the cue-out point, \tablename~\ref{tab:policy} shows an opposite result. That is, in terms of the number of cue points closest to the human annotations, the cue-out point is the second and the cue-mid point is the worst among the three types. These results indicate that the cue-mid point is a safe choice. That is, although the cue-mid point is least likely to be a boundary as shown in  \tablename~\ref{tab:policy}, the difference between the estimate and human annotation is relatively small because it is the middle of the transition region. 


\tablename~\ref{tab:hit} shows the difference between human-annotated boundary timestamps and one of the cue point types in terms of median time (in seconds) on the left side. The overall trend confirms that the cue-in is the best estimate of track boundary and the cue-mid is a safer choice than the cue-out. The table also shows the result of ``cue-best''. This is computed with the minimum difference among the three cue point types for each of the transition region. The result shows that the median time differences are dramatically decreased to 4-5 seconds. \tablename~\ref{tab:hit} also shows the difference between human-annotated boundary timestamps and the cue-in point in terms of hit rates on the right side. The hit rates are computed the ratio of correct estimates given a tolerance window. If the estimate is within the tolerance window on the human-annotated boundary timestamp, it is regarded as a correct estimate. We set three tolerance windows (15, 30, and 60 seconds) considering that the average tempo of tracks in the dataset is 127 beat per minute (BPM) and then the tolerance windows approximately correspond to 32, 64, 128 beats (multiples of a phrase unit). The result shows that the best hit rate with the 30 second window (about 64 beats) is above 80\%. Given the long transition time as shown in \figurename~\ref{fig:cue_translength}, the cue-in point may be considered as a reasonable choice.

\subsection{Effect of Audio Features}

\tablename~\ref{tab:hit} also compares the median time difference between human-annotated boundary timestamps and one of the cue point types for different audio features used in the subsequence DTW. In general, the chroma features are a better choices than MFCC (p-value of t-test < 0.001 for chroma with or without key-invariant). When both of chroma and MFCC are combined, the median time difference slightly reduces but it is statistically insignificant (p-value of t-test > 0.1). However, we observed that the subsequence DTW does not work well for some genres such as Techno which only contain drum and ambient sounds. This might can be improved by using MFCCs with a large number of bins or using mel-spectrograms. The use of key-invariant chroma generally does not make much difference because key transposition does not performed frequently as discussed in Section~\ref{sec:key_trans}.



\section{Musicological Analysis of DJ Mixes}
\label{sec:tempokey_analysis}
We hypothesize that DJs share common practices in the creative process in terms of tempo change, track-to-track transition, and cue point selection. In this section, we validate them using the results from the mix-to-track subsequence alignment and the cue point extraction.

\subsection{Tempo Adjustment}

\begin{figure}[t]
    \centering
    \hspace{-1cm}
    \includegraphics[width=.8\columnwidth]{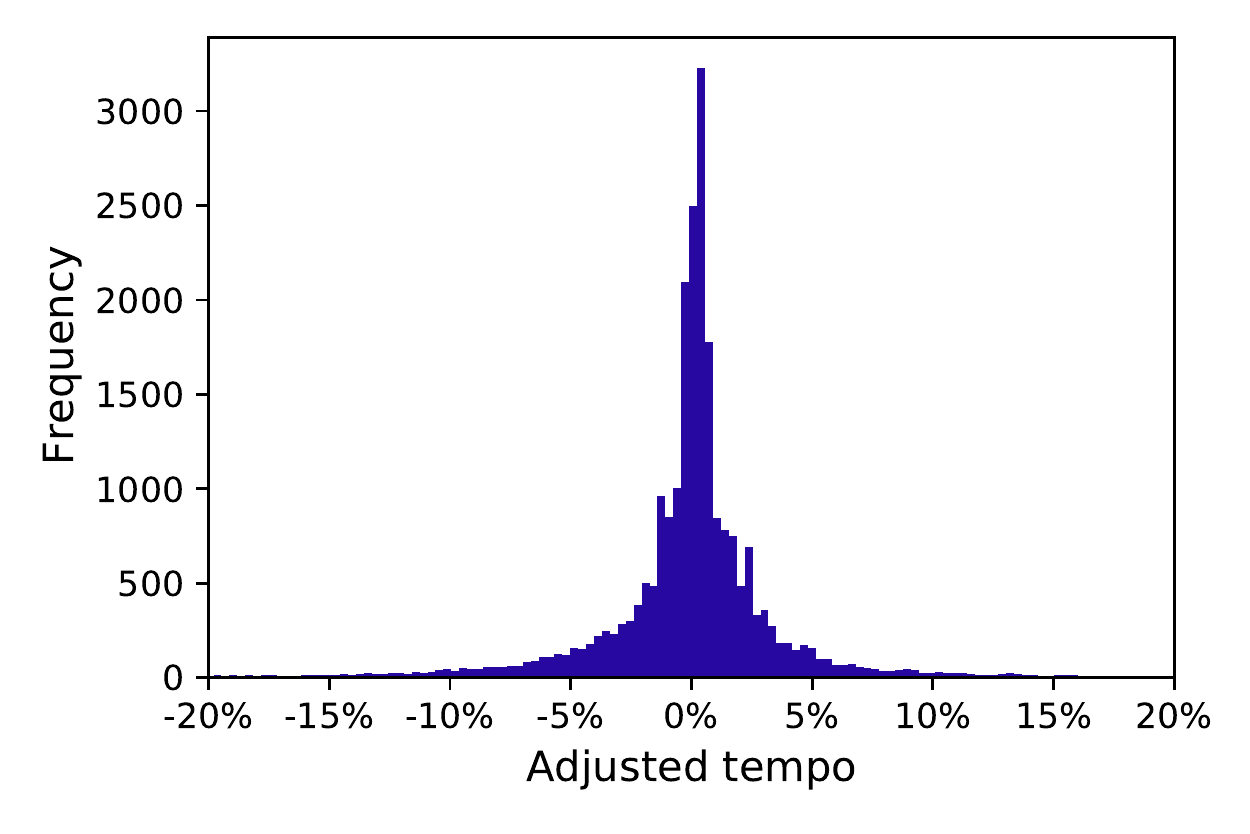}
    \vspace{-5mm}
    \caption{A histogram of adjusted tempo of tracks in mixes.}
    \label{fig:diff_bpm}
\end{figure}

We compare the estimated tempo of the original track to the tempo of each audio segment where the track is played in a mix. \figurename~\ref{fig:diff_bpm} shows a histogram of percentage differences of tempo between the original track and the audio segment in the mix. For example, a difference of 5\% indicates the tempo of the original track is increased by 5\% while played in the mix. As shown in the histogram, the adjusted tempo has an double exponential distribution, which means the adjusted tempo values are skewed towards zero. In detail, 86.1\% of the tempo are adjusted less than 5\%, 94.5\% are less than 10\%, and 98.6\% are less than 20\%. If one implements an track identification system for DJ mix that is robust to tempo adjustment, this distribution could be a reference.  

\subsection{Key Transposition}
\label{sec:key_trans}
\begin{figure}[!t]
    \centering
    \hspace{-1.1cm}
    \includegraphics[width=.65\columnwidth]{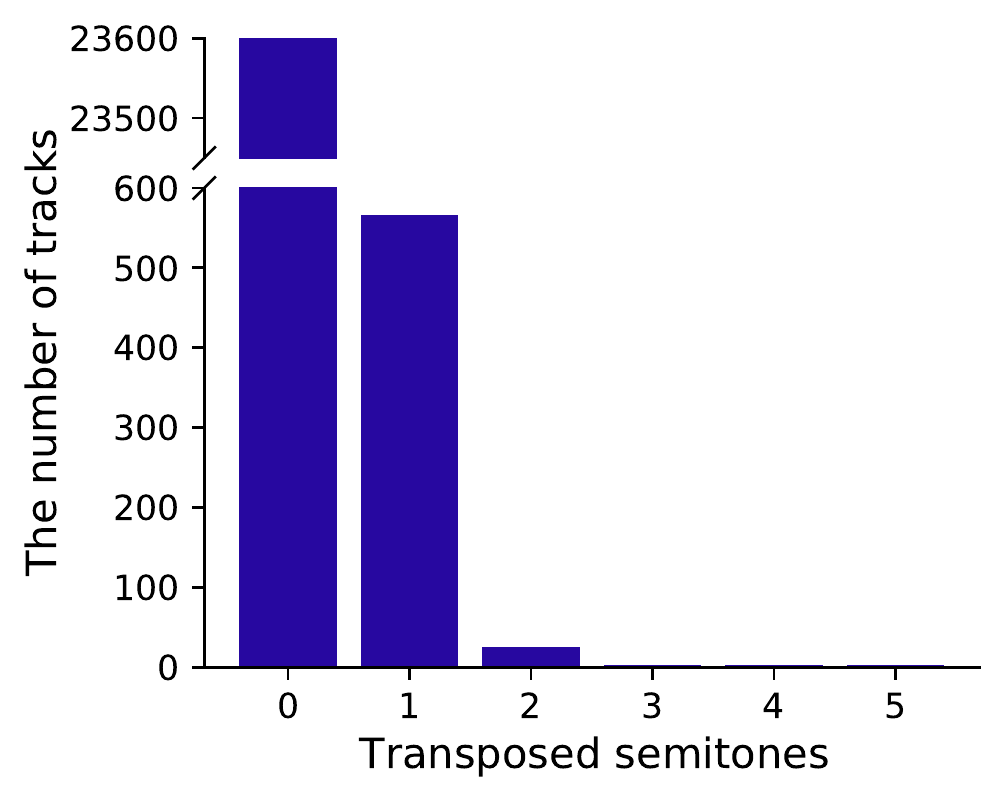}
    \vspace{-3.8mm}
    \caption{The number of tracks depending on the number of semitones in mixes.}
    \label{fig:diff_key}
\end{figure}

A function so-called ``master tempo'' or ``key lock'' that preserves pitch despite tempo adjustments is activated by default in modern DJ systems such as stand-alone DJ systems, DJ softwares, and even turntables for vinyl records.
Therefore, key transposition is usually performed when a DJ intentionally wants to change the key of a track.
As mentioned in Section~\ref{sec:dtw}, the transposition-invariant DTW can provide the number of transposed semitones as a by-product. We computed the statistics of key transposition using them (using DTW taking both MFCCs and key-invariant chroma). \figurename~\ref{fig:diff_key} shows a histogram of key transposition between the original track and the audio segment in the mix. Only 2.5\% among the total 24,202 tracks are transposed and, among those transposed tracks, 94.3\% of them are only one semitone transposed. This result indicates that DJs generally do not perform key transposition much and leave the ``master tempo'' function turned on in most cases.



\begin{figure}[t]
    \centering
    \hspace{-.3cm}
    \includegraphics[width=\columnwidth]{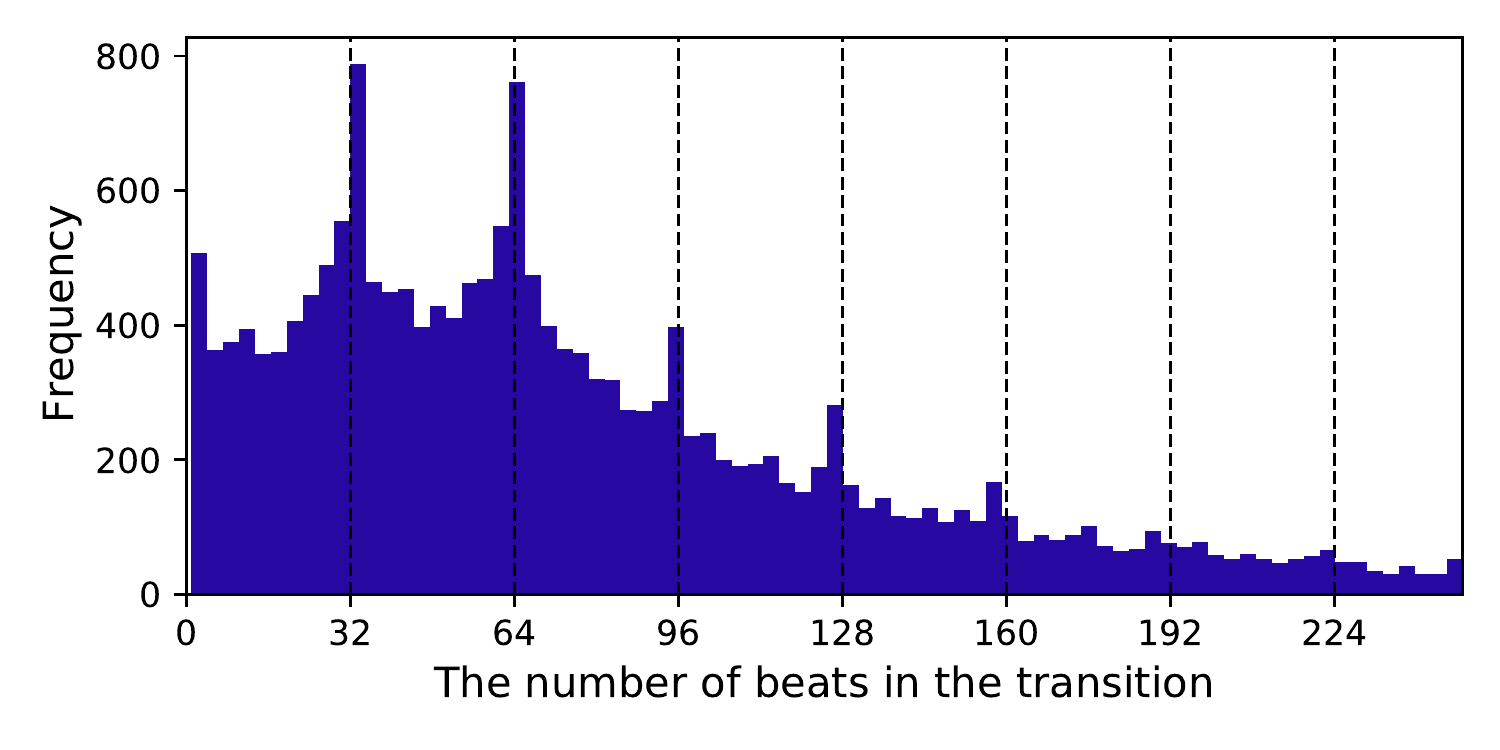}
    \vspace{-5mm}
    \caption{
    A histogram of the transition lengths in number of beats.
    The dotted lines are plotted at every 32 beats.
    }
    \label{fig:cue_translength}
\end{figure}

\begin{figure}[t]
  \centering
  \hspace{-.5cm}
  \includegraphics[width=.81\columnwidth]{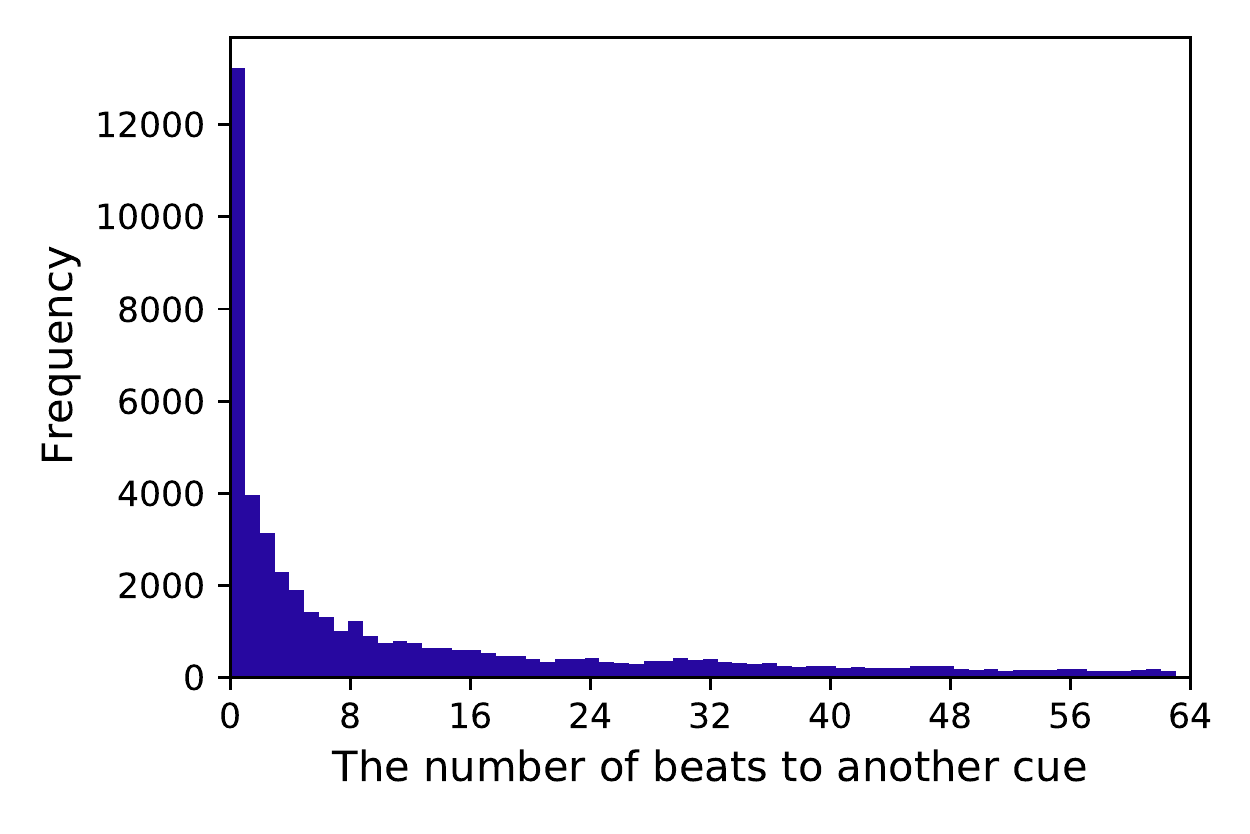}
  \vspace{-5mm}
  \caption{
    A histogram of distances between cue points of a single track in the number of beats. 
  }
  \label{fig:cue_diff}
\end{figure}

\subsection{Transition Length}
\label{sec:trans_length}
Once we extract cue-in and cue-out points in the transition region, we can calculate the transition length. This can provide some basic hints on how DJ makes the track-to-track transition in a mix. \figurename~\ref{fig:cue_translength} shows a histogram of transition lengths in the number of beats. We annotated the dotted lines every 32 beat which is often considered as a phrase in the context of dance music. The histogram has peaks at every phrase. This indicates that DJs consider the repetitive structures in the dominant genres of music when they make transitions or set cue points.
 

\subsection{Cue Point Agreement among DJs}
Deciding cue points of played tracks is a creative choice in DJ mixing. Observing the agreement of cue points on a single track among DJs may elucidate the possibility of finding some common rules. To the end, we collected all extracted cue points for each track and computed the statistics of deviations in cue-in points and cue-out points among DJs. Specifically, we computed all possible pairs and their distances separately for cue-in points and cue-out points. Since the two distributions were almost equal, we combined them into a single distribution in \figurename~\ref{fig:cue_diff}. From the results, 23.6\% of the total cue point pairs have zero deviation. 40.4\% of them were within one measure (4 beats), 73.6\% were within 8 measures and 86.2\% were within 16 measures. This indicates that there are some rules that DJs share in deciding the cue points. It would be interesting to perform detailed pattern analysis to estimate the cue points using this data in future work.


\vspace{-2mm}
\section{Conclusions}
We presented various statistics and analysis of 1,557 real-world DJ mixes from \emph{1001Tracklists}.
Based on the mix-to-track subsequence DTW, we conducted cue point analysis of individual tracks in the mixes and showed the possibility of common rules in the music making that DJs share.
We also investigated mix segmentation by comparing the three types of cue point to human-annotated boundary timestamps and showed that humans tend to recognize cue-in points of the next tracks as boundaries. Finally, we showed the statistics of tempo and key changes of the original tracks in DJ performances. We believe this large-scale statistical analysis of DJ mixes can be beneficial for computer-based research on DJ music. The cue point analysis can be the ground for the precise definition of cue points and the tempo and key analysis can provide a guideline of the musical changes during the DJ mixing. 



As a future work, we plan to estimate cue points within a track as a step towards automatically generating a mix~\cite{schwarz2018heuristic,veire2018raw}. The cue point estimation has many application such as DJ software and playlist generation on music streaming services. This will require structure analysis or segmentation of a single music track, which is an important topic in MIR.  
Furthermore, we plan to analyze the transition region in a mix to investigate DJ mixing techniques. For example, it is possible to estimate the gain changes in the cross-faded region by comparing the two adjacent original tracks and the mix~\cite{schwarz2019methods,werthen2018ground}.
The methods can be extended to the spectrum domain. Such detailed analysis of mixing techniques will allow us to understand how DJs seamlessly concatenate music tracks and provide a guide to develop automatic DJ systems.   




\section{Acknowledgement}
We greatly appreciate \emph{1001Tracklists} for offering us the mix metadata  employed in this study.
We note that the metadata used for this analysis was obtained with permission from \emph{1001Tracklists}, and suggest that people who are interested in the data contact \emph{1001Tracklists} directly.
This research was supported by BK21 Plus Postgraduate Organization for Content Science (or BK21 Plus Program), Basic Science Research Program through the National Research Foundation of Korea (NRF-2019R1F1A1062908), and a grant from the Ministry of Science and Technology, Taiwan (MOST107-2221-E-001-013-MY2).


\bibliography{ref}

\end{document}